\documentclass[prb,twocolumn,showpacs,preprintnumbers,amsmath,amssymb,showkeys]{revtex4}

\usepackage{amsmath}
\usepackage{amssymb}
\usepackage{bm}
\usepackage{graphicx}
\usepackage{subfigure}

\begin{document}

\title{Transition metal dimers as potential molecular magnets: A challenge to computational chemistry}

\author{Daniel Fritsch}

\affiliation{Leibniz Institute for Solid State and Materials Research, IFW Dresden, PO Box 270116, D-01171 Dresden, Germany}

\author{Klaus Koepernik}

\affiliation{Leibniz Institute for Solid State and Materials Research, IFW Dresden, PO Box 270116, D-01171 Dresden, Germany}

\author{Manuel Richter}

\affiliation{Leibniz Institute for Solid State and Materials Research, IFW Dresden, PO Box 270116, D-01171 Dresden, Germany}

\author{Helmut Eschrig}

\affiliation{Leibniz Institute for Solid State and Materials Research, IFW Dresden, PO Box 270116, D-01171 Dresden, Germany}

\date{\today}

\begin{abstract}
Dimers are the smallest chemical objects that show magnetic anisotropy. We focus on 3$d$ and 4$d$ transition metal dimers that have magnetic ground states in most cases. Some of these magnetic dimers have a considerable barrier against re-orientation of their magnetization, the so-called magnetic anisotropy energy, MAE. The height of this barrier is important for technological applications, as it determines, e.g., the stability of information stored in magnetic memory devices. It can be estimated by means of relativistic density functional calculations. Our approach is based on a full-potential local-orbital method (FPLO) in a four-component Dirac-Kohn-Sham implementation. Orbital polarization corrections to the local density approximation are employed. They are discussed in the broader context of orbital dependent density functionals. Ground state properties (spin multiplicity, bond length, harmonic vibrational frequency, spin- and orbital magnetic moment, and MAE) of the 3$d$ and 4$d$ transition metal dimers are evaluated and compared with available experimental and theoretical data. We find exceptionally high values of MAE, close to 0.2 eV, for four particular dimers: Fe$_2$, Co$_2$, Ni$_2$, and Rh$_2$.
\end{abstract}

\pacs{31.15.E-, 31.15.aj, 75.30.Gw, 75.50.Xx}

\keywords{density functional calculations, 3$d$ and 4$d$ transition metal dimers, magnetic anisotropy energy, MAE, orbital polarization correction, OPC}

\maketitle

\section{Introduction}
\label{Introduction}

During the past decades we have been eyewitnesses of enormous
progress in experimental material design.
It has become common that this experimental progress
is accompanied by material-specific theoretical description.
Beyond this, theoretical modelling has even partly replaced 
expansive experimental scannings of large series of 
samples with, e.g., different compositions.
This advance has become possible due to the large predictive power
of quantum mechanical calculations, in particular of density
functional approaches.

Of particular scientific and practical interest is the study
of small particles. In the realm of chemistry, the smallest
considered objects are dimers. While many main group elements form natural
homonuclear dimers at ambient conditions, the production and experimental study
of metal dimers is difficult. On the other hand, as will be demonstrated
in the present work, some metal dimers show exceptional magnetic
properties that are of potential interest for, e.g., magnetic
information storage technology.
It is a major challenge to computational chemistry to identify
the most promising homo- and heteronuclear dimers
that deserve intense experimental
study and further theoretical modelling, e.g., related to a possible
embedding into an appropriate matrix.
In the present study, we will concentrate on homonuclear dimers
of 3$d$ and 4$d$ transition metal elements.

While spin magnetism is routinely considered in spin density
functional calculations, the orbital contribution to the
magnetization is frequently neglected. 
This has two reasons: (i) in systems with high coordination
like large clusters or bulk, the orbital moment is almost
quenched due to chemical bonding; thus, the magnetization 
almost exclusively originates from the spins;
(ii) technically, orbital magnetic properties require
much more (at least about one order of magnitude) computational
effort, as spin-orbit effects have to be considered.
On the other hand, it is spin-orbit interaction that
brings about one of the most important properties of
magnetic systems: the magnetic anisotropy (MA).
At low temperature, a given system resides in its magnetic ground
state if the magnetization directs along the
so-called easy axis of the system.
The magnetic anisotropy energy (MAE) 
is the energy amount which would be needed to rearrange the
magnetic moments of the system in the direction of the so-called hard axis.
(Depending on the symmetry, several hard axes or hard planes
may exist.)
Evidently, the larger the MAE the more stable the
system behaves against magnetic moment
flips and a possible information stored in this way is more safe. Thus, systems with large MAE are interesting for technological application, 
e.g., in magnetic storage devices.

Due to enlarged computational resources and more advanced
numerical routines, density functional methods are nowadays
capable of reliably determine total energy differences well below
mHartree accuracy in many cases.
This accuracy is at least required when calculating MAE,
which are small total energy differences.
A scan through the existing literature reveals that 
the magnitude of MAE depends on the dimensionality
of the system under consideration. As
an illustrative example we refer to the estimates for
Co systems of different dimensionalities as given by Gambardella
\textit{et al.}\cite{Gambardella_Nature416_301}
Therein, the MAE is estimated to increase by roughly 
two orders of magnitude when decreasing the dimensionality of the Co system.
Amounting to 0.04 meV (per atom) for bulk hcp Co, the MAE increases to
0.14 meV (per atom) for a monolayer of Co atoms, and increases further
to 2.0 meV (per atom) for a chain of Co atoms with averaged length of 80 atoms.
Now, the most interesting question is, how large the MAE of a transition
metal system of even smaller size might be.
This leads us directly to the 3$d$ and 4$d$ transition metal dimers,
being among the smallest systems showing anisotropy effects.
We will demonstrate that potentially the MAE can be {\em two further
orders of magnitude higher} than discussed by Gambardella \textit{et al.}\cite{Gambardella_Nature416_301}

Substantial experimental work has been carried out on transition metal dimers.
It has been summarized in a recent review by Lombardi and Davis.\cite{Lombardi_ChemRev102_2431}
There is also an older review of Morse,\cite{Morse_ChemRev86_1049} summarizing the experimental data up to 1986.
It has been mentioned by several authors that the computational investigation of dimers,
though seemingly the most simple chemical system, is far from being trivial.
This holds in particular for transition metal dimers, where frequently a number of
low-lying states of different spin multiplicity and symmetry compete.
There are several works published on the theoretical investigation
of 3$d$ and 4$d$ transition metal dimers, mainly focusing on the structural properties. 
The 3$d$ transition metal dimers have recently been investigated by 
Yanagisawa \textit{et al.}\cite{Yanagisawa_JChemPhys112_545} using a hybrid DFT method.
In the same year,
Barden \textit{et al.}\cite{Barden_JChemPhys113_690} performed calculations
using six density functional or hybrid Hartree-Fock/density functional theory methods
and an unrestricted Hartree-Fock method.
Later, Gutsev \textit{et al.}\cite{Gutsev_JPhysChemA107_4755} reported
on calculations using several approximations to the density functional 
as implemented in the Gaussian 98 program package,
whereas Valiev \textit{et al.}\cite{Valiev_JChemPhys119_5955} applied
the projected augmented plane wave method to calculate structural
properties of the 3$d$ transition metal dimers.

Continuing with the 4$d$ transition metal dimers, we are aware of only
two works on their structural properties.
In an early work, Wu\cite{Wu_ChemPhysLett383_251} investigated
the 4$d$ transition metal dimers from Y$_2$ to Cd$_2$ except Tc$_2$ utilizing
several approximations to the density functional as implemented 
in the Gaussian 98 suite of programs.
More recently, Yanagisawa \textit{et al.}\cite{Yanagisawa_JCompChem22_1995}
reported on their investigation on 4$d$ transition metal dimers from Y$_2$ to Ag$_2$.
They again applied several approximations to the density functional.
Additionally, they carried out calculations
using M\o{}ller-Plesset second-order perturbation method for comparison.

All the abovementioned work is restricted to the investigation of 
structural properties, such as bond lengths, harmonic vibrational
frequencies, and dissociation energies. 
We are aware of only two, very recent publications devoted to
the calculation of orbital moments and MAE for transition metal
dimers. 
Strandberg \textit{et al.}\cite{Strandberg_NatureMaterials6_648}
published data on the MAE of Co$_2$, Ni$_2$, and Rh$_2$,
and Fern\'andez-Seivane and Ferrer reported related data for
Pd$_2$, Ir$_2$, Pt$_2$, and Au$_2$.\cite{Fernandez-Seivane_prl99_183401}
Both publications used local spin density approximation (LSDA) and/or
generalized gradient approximation (GGA) to the density functional.
We are not aware of any MAE calculation for a complete 3$d$ or 4$d$
series of homonuclear transition metal dimers. Further, we are
not aware of any calculation for the MAE of dimers beyond
local approaches to the density functional. Both
topics will be addressed in the present study.
The presented results might become an issue of technological relevance
once one would be able to fix the position and geometrical orientation
of dimers without much disturbing their magnetic state.

In the following Section~\ref{OrbDepPotDFT}, we will briefly
review the necessary theoretical background for the use of orbital
dependent potentials within density functional theory.
The treatment of the magnetic anisotropy in general and the need
to introduce so-called orbital polarization corrections (OPC)
is discussed in Section~\ref{MagneticAnisotropy}, 
whereas Section~\ref{ComputationalDetails} summarizes the computational details.
In Section~\ref{StructuralProperties},
the results on ground state spin multiplicity,
bond lengths and harmonic vibrational frequencies
are presented and discussed.
Section~\ref{MagneticProperties} compiles results on
the magnetic properties (spin and orbital magnetic moments and MAE),
calculated without and with orbital polarization corrections.
Finally, Section~\ref{SummaryAndOutlook} summarizes the findings
of the present work and gives an outlook.

\section{Orbital dependent potentials in DFT}
\label{OrbDepPotDFT}

Density functional theory (DFT) has proved an enormous predictive power for real materials, although the model density functionals in use (LDA, GGA, B3LYP, \ldots, just to mention a few acronyms) are partially based on uncontrolled model assumptions.  Nevertheless, further progress of DFT depends on designing better model density functionals for chosen fields of application. (Although the existence of a universal density functional has been proven, there is little hope to model this probably extremely complex functional in general.) In recent time, more and more orbital dependent functionals like self-interaction correction (SIC), exact exchange (EXX), local spin density approximation plus local correlation (LSDA+$U$), OPC and others came into the game. (Local density approximation plus dynamical mean
field theory (LDA+DMFT) is a combined density-functional plus other-many-body-techniques approach to the self-energy of the single-particle Green's function and is another important development aiming mainly at electronic excitations.) Orbital dependence means non-locality of the potential operator (integral operator), and this costs appreciable additional computing effort. Therefore, local approximants of these potentials like optimized effective potentials (OEP) are popular. Sometimes it is also incorrectly claimed that density functional methods would demand local effective potentials. True, if the universal density functional $H[n]$ has a unique functional derivative, this corresponds to a local \emph{external} potential
$(v(\bm r)-\mu) = -\delta H/\delta n(\bm r)$ where $\mu$ is the chemical potential. This results from the Hohenberg-Kohn variational principle
\begin{equation}
  \label{eq:II.1}
  E[v,N] = \min_n \bigl\{H[n] + (v\,|\,n) \,\bigl|\, (1\,|\,n) = N\bigr\},
\end{equation}
where
\begin{equation}
  \label{eq:II.2}
  (v\,|\,n) = \sum_{ss'}\int d^3r v_{ss'}(\bm r)n_{s's}(\bm r)
\end{equation}
and the spin density components vary in $L^3(V)$ of cubic summable functions on the position space $V$ of finite volume.\cite{e171} Cubic summability comes from the demand of finiteness of kinetic energy.\cite{lieb83} The existence and uniqueness of the functional derivative of $H[n]$ \emph{on hypersurfaces of constant particle number $N$ or particle numbers $N_\sigma$ of the spin channels of non-relativistic theory} is the statement of the first theorem or lemma by Hohenberg and Kohn (adapted to a whole functional space for the variation, comprising also cases of degenerate ground states). It is equally well known that this derivative jumps across those hypersurfaces.

To perform the variation, one may introduce the generalized Kohn-Sham ansatz
\begin{widetext}
\begin{equation}
  \label{eq:II.3}
  H[n] = K[n] + L[n], \quad
  K[n] = \min_{\phi_k,n_k}\bigl\{{\cal K}[\phi_k,n_k] \,\bigl|\,
  (\phi n \phi) = n,\; \langle\phi_k\,|\,\phi_l\rangle = \delta_{kl},\; 
  0 \leq n_k \leq 1\bigr\}
\end{equation}
\end{widetext}
with the abbreviation
\begin{equation}
  \label{eq:II.4}
  n = n_{ss'}(\bm r) = \sum_k \phi_k(\bm rs) n_k \phi^*_k(\bm rs') 
  \equiv (\phi n \phi)
\end{equation}
and where ${\cal K}[\phi_k,n_k]$ is an explicitly given functional having a unique minimum with respect to variation of the orbitals $\phi_k(\bm rs)$ and their occupation numbers $n_k$ under the indicated
constraints, while $L = H - K$ further depends on (or determines) the level of knowledge of the total density functional $H[n]$.

Then, the Kohn-Sham variational principle
\begin{widetext}
\begin{equation}
  \label{eq:II.5}
  E[v,N] = \min_{\phi_k,n_k} \biggl\{{\cal K}[\phi_k,n_k] + L[(\phi n \phi)] + 
  \sum_{kl}\langle\phi_l|\sqrt{n_l}(v\delta_{lk} - \varepsilon_{lk})
  \sqrt{n_k}|\phi_k\rangle \,\biggl|\, 
  0 \leq n_k \leq 1,\; \sum n_k = N\biggr\}
\end{equation}
\end{widetext}
is obtained. The $\varepsilon_{lk}$ are Lagrange parameters for the orthonormality constraints.  If as usually $\cal K$ includes an kinetic energy term of orbitals, then the minimal density $n$ from (\ref{eq:II.5}) is automatically in $L^3(V)$. With the writing
\begin{equation}
  \label{eq:II.6}
  \frac{1}{n_k}\,\frac{\delta{\cal K}}{\delta\phi^*_k} = 
  \frac{f_k[\phi_l,n_l]}{n_k} \equiv \hat k_k|\phi_k\rangle, \quad
  \frac{\delta L}{\delta n} = v_\textrm{L}
\end{equation}
the variation leads to the generalized Kohn-Sham equations
\begin{equation}
  \label{eq:II.7}
  \langle\phi_l|\hat k_k + v_{\text{L}} + v|\phi_k\rangle = 
  \sqrt{n_l/n_k}\varepsilon_{lk}
\end{equation}
and to the ordinary aufbau principle for the occupation numbers $n_k$. In general they may be solved together with the orthonormality constraints by a unitary triangulation of the matrix of t he left hand side expressed in a suitable order of an orthonormal orbital basis. If as in many cases $\hat k_k$ is a Hermitian operator, a diagonalization is possible.

In the ordinary Kohn-Sham approach $\hat k_k$ consists of the orbital kinetic energy operator which is the same differential operator for all orbitals. (The classical Coulomb potential of the density $n$ can
likewise be considered as an orbital expression or as part of $v_{\text{L}}$, here we adopt the latter alternative.) Nevertheless, discontinuities of the functional derivative of $K[n]$ when varying the
particle number are incorporated in this approach. They yield a piecewise linear behavior of this contribution to
the total energy as a function of particle number and an appreciable part of band gap energies. Improvement of ionization
potentials, electron affinities and band gaps can be expected from further orbital related terms introduced in $\hat k_k$ as is the case in the SIC, LSDA+$U$, EXX, OPC and other approaches. In particular in magnetism, the non-relativistic total energy may be piecewise linear even with the total particle number fixed, if it changes between the two spin channels as analyzed in Ref.~\onlinecite{e154}. Note that every piecewise linear behavior of the total energy functional causes a discontinuity in the derivative of its Legendre transform which is the density functional. Therefore, there may be good reasons not to remove the
orbital dependence of potentials from model functionals.

The numerical results in the present paper are obtained with the LSDA and the LSDA+OPC functionals. Since we are aiming at magnetic anisotropy energy which is due to spin-orbit coupling, the Dirac-Kohn-Sham
equations are solved. Then, in the LSDA case, $\hat k_k$ reduces to the Hermitian relativistic kinetic energy operator $\hat t$ and the orbitals are four-component bispinor orbitals. In the case of LSDA+OPC one has
\begin{equation}
  \label{eq:II.8}
  \hat k_k = \hat t + \hat v_{\text{OPC}},
\end{equation}
with the Hermitian non-local potential
\begin{equation}
  \label{eq:II.9}
  \hat v_{\text{OPC}} = \sum_{\nu,\nu'}^{(\bm R_\nu = \bm R_{\nu'})}
  |\nu\} C_{\nu\nu'} \{\nu'|.
\end{equation}
The multi-index $\nu$ stands for $\nu = (\bm R,\rho,\kappa,\mu)$, where $\bm R$ is the atomic position, $\rho$ is the principal atomic site orbital quantum number, $\kappa$ is the spin-orbit quantum number linked
to the orbital angular momentum quantum number by $l(\kappa) = \kappa$ for $\kappa>0$ and $l(\kappa) = -\kappa-1$ for $\kappa<0$ in the absence of external magnetic field (in the presence of a magnetic field this connection still holds to a very good approximation for the large bispinor component), and $\mu$ is the $z$-component of the total angular momentum with the $z$-axis chosen in direction of the site magnetic polarization.\cite{e180} The non-orthogonal four-component basis orbitals for the algebraization of the Dirac-Kohn-Sham equation used in 
the full-potential local-orbital (FPLO) package\cite{Koepernik_prb59_1743,FPLO}
are denoted by $|\nu\rangle$, and $|\nu\}$ are the states of the contragradiant basis, $\{\nu|\nu'\rangle = \delta_{\nu\nu'}$. The matrix elements of the used OPC version\cite{Nordstrom92} are
\begin{equation}
  \label{eq:II.10}
  C_{\nu\nu'} = B_{l(\kappa)} M_{\sigma} \langle\nu|\hat L_z|\nu'\rangle\; ,
\end{equation}
where $\hat L_z$ is the operator of the $z$-component of the orbital angular momentum,
\begin{equation}
  \label{eq:II.11}
  M_{\sigma} = \sum_k n_k\langle\phi_k|P_{\sigma}|\nu\}
               \langle\nu|\hat L_z|\nu'\rangle
               \{\nu'|P_{\sigma}|\phi_k\rangle
\end{equation}
is the $z$-component of the obtained orbital momentum in the considered atomic spin-subshell ($P_{\sigma}$ being projection operators to the spin channel),
 and $B_l$ is a Racah coefficient ($B_2 = (9F^2 -5F^4)/441$ in terms of Slater integrals with
the related 3$d$ or 4$d$ radial functions).
 The $\nu$ and $\nu'$ sums in (\ref{eq:II.9}) run over the `magnetic' basis orbitals which are
the 3$d$ or 4$d$ basis orbitals in our case being quite local and well representing the atomic $d$ shells of the atoms in the dimer (or in other cases in a solid).

\section{Magnetic anisotropy}
\label{MagneticAnisotropy}

Magnetic anisotropy is the dependence of measurable quantities on the direction
of the applied magnetic field, $\bm{B}$. In a more narrow sense, but much more convenient for
theoretical considerations, one can consider MA as the dependence of free or inner energy on the direction of magnetization, $\bm{M}$.

In extended systems, both viewpoints are commonly related by splitting\cite{Kuzmin_encyc} the generating function of free energy, $\Phi (T,\bm{B},\bm{M} )$:
\begin{equation}
\Phi = \Phi_0 + E_{\rm MA}(\bm{e}) - |\bm{M}| \bm{e} \bm{B} \; .
\end{equation}
Here, the dependence on the direction vector of magnetization, $\bm{e} = \bm{M}/|\bm{M}|$, and on applied magnetic field has been made explicit, but the first two terms also depend on $|\bm{M}|$.
They further depend on temperature $T$, pressure, {\em etc}. Note, that the subdivision into the isotropic part $\Phi_0$ and the anisotropic part $E_{\rm MA}$
is not unique and that the free energy is the minimum of $\Phi$ with respect to $\bm{M}$.

The suggested notation makes sense if the field-induced change of $(\Phi_0 + E_{\rm MA}(\bm{e}))$ via variation of $|\bm{M}|$ is much weaker than the change of $E_{\rm MA}(\bm{e})$ via variation of $\bm{e}$. This is usually the case, e.g., for ferromagnets well below the magnetic ordering temperature in moderate applied fields.\cite{Kuzmin_encyc}

For the present purpose, we will adopt a somewhat different notation. Considering finite systems,
thermodynamic features can be dropped. The focus is now on the dependence of ground state energy, $E_0$, on the direction of magnetization, to be evaluated by constrained DFT. Define
\begin{equation}
{\rm MAE}(\bm{e}) = E_0 (\bm{e}) -  E_0 (0, 0, 1) \; ,
\label{EQ:MAE}
\end{equation}
where $(0,0,1)$ denotes the direction parallel to some appropriately
chosen axis, e.g., the dimer axis.
This definition does not rely on a weak dependence of the anisotropic part of the energy on the magnitude of the total moment, since
the relation with previous notation is $E_0 ({\bm e}) = E_{\rm MA}(\bm{e}, |{\bm M}|(\bm{e}), T\rightarrow 0)$
for appropriate choice of $\Phi_0$.
As we will see in the results, the ground state value of the total
magnetic moment,
\begin{equation}
\int d^3r \;|\bm{M}| = |\langle 2 \hat{\bm{S}} + \hat{\bm{L}}\rangle | \mu_{\rm B} = \mu_s + \mu_l
\end{equation}
may indeed substantially depend on $\bm{e}$.
This direction is imposed on the system by constraining the xc field parallel to chosen $\bm{e}$. As only collinear spin- and magnetization densities are considered,
\begin{equation}
\bm{\mu}_s + \bm{\mu}_l = \bm{e} \langle 2 \hat{S_z} + \hat{L_z} \rangle \mu_{\rm B} \; .
\end{equation}
Note, that $\mu_l = \mu_{\rm B} \sum_{\sigma} M_{\sigma}$ relates the
notation in terms of magnetic quantum numbers, Eq.~(\ref{eq:II.11}),
with that in terms of orbital moments.

Dimers are the smallest chemical objects that show MA.\cite{Strandberg_NatureMaterials6_648}
While $E_0 (\bm{e})$ of
atoms without external magnetic field is isotropic and obtains
axial symmetry only by application of a field, dimers possess axial symmetry at $B=0$. One could argue that free dimers would always orient themselves in the external magnetic field such that the lowest possible energy is achieved. This statement is, however, valid for any finite system.
For example, macroscopic samples are frequently positioned free to rotate in magnetic
measurements in order to ensure orientation of $\bm{M}$ along $\bm{B}$.\cite{Kuzmin04} What we aim at, however, is to investigate dimers with a rotation axis fixed in space by an unspecified interaction, e.g., as a model for dimers on a substrate. This model interaction should be strong enough to withstand the torque due to the magnetic field. On the other hand, it should not essentially influence the electronic structure of the dimer and, thus, the MAE defined in Eq.~\eqref{EQ:MAE}.

One challenge to both computational and applied chemistry would be to find an embedding medium fullfilling the two opposing requirements in order to verify the enormous MAE values predicted by our calculations and two preceding papers.\cite{Strandberg_NatureMaterials6_648,Fernandez-Seivane_prl99_183401} If such a medium was found, the MAE values would represent barrier heights against thermal and quantum mechanical fluctuations of the super-paramagnetic dimer moments. A 10-year stability of super-paramagnetic bits would require MAE/$kT > 40$, $k$ being the Boltzmann constant.\cite{ref40kt} Our results will show that such a stability could be achieved for $T \approx 60$ K. Thus, the search for dimers (or other very small clusters) with room-temperature stability should not be considered an unrealistic goal.

In a non-relativistic theory, spin and real space are independent of each other and spin magnetization can be rotated by the external field without change of energy. Thus, the consideration of spin-orbit coupling (and other relativistic effects on the level of the Dirac-Kohn-Sham formalism) is needed for the description of MAE. Here, a four-component relativistic approach is used.\cite{e180} Beyond these kinematic relativistic effects, magnetic dipole-dipole interaction contributes to MAE. This interaction is the source of so-called shape anisotropy in macroscopic theory. It is responsible for the fact that iron magnets should have special shape,
e.g., that of horse shoes or long bars, in order to perform their operation. In formal theory, this interaction is described as Hartree-like approximation to the Breit interaction of relativistic quantum electrodynamics.\cite{Jansen88} Here, we neglect the shape anisotropy as it is small (about 1 meV) compared with the spin-orbit generated anisotropy in most of the considered dimers.

Orbital dependent exchange effects to the MAE of dimers are more important in some cases and have been neglected in the previous approaches.\cite{Strandberg_NatureMaterials6_648,Fernandez-Seivane_prl99_183401}
Such effects are not included in the standard DFT approximations like LSDA or GGA, which are based on homogeneous electron gas theory. For example, LSDA does not account for the second Hund rule (maximum orbital quantum number in the ground state)
in atoms.\cite{Eschrig_EuroPhysLett72_611} Thus, it is likely that also the ground state of dimers is not accurately enough described for the present purpose in any known local approximation. 
For this reason, we have decided to add orbital polarization corrections (OPC) to LSDA, as described
in the previous section, in the present calculations.
Such corrections were first suggested by Brooks, Eriksson, and Johansson.\cite{Brooks85,Eriksson90} The specific implementation used is taken from Ref. \onlinecite{Nordstrom92}.
Though this is just one out of several similar approaches it will provide an impression how strong the influence of orbital exchange terms on such a sensitive quantity like MAE can be. As a matter of (scarce) experience, LSDA gives a lower bound on the MAE values while LSDA+OPC gives an upper bound, at least for systems with low coordination. An example where both experimental and theoretical data are available is Co atoms and small Co clusters on the surface of Pt. Here, the strength of OPC had to be down-scaled by 50\% in order to meet the experimental values of MAE.\cite{Gambardella_Science300_1130} Below, we will present and discuss both LSDA and LSDA+OPC data. This will allow the reader to get an idea of the order of magnitude that can be expected in reality.

While DFT approaches to (orbital) magnetic properties of extended systems 
are state-of-the-art, considerable less experience is available for small
clusters. We thus feel inclined to provide the following arguments to justify our approach:
\begin{itemize}
\item Spin magnetism and spin-orbit coupling in the ground state of single atoms and of weakly correlated solids are very well described in LSDA.\cite{Richter01}
\item Orbital magnetism in the ground state of single atoms is by construction well described in LSDA+OPC; it works reasonably well in metallic magnets.\cite{Richter01,Eschrig_EuroPhysLett72_611}
\item There are cases (e.g., lanthanides) where the ground state configuration and ionization
potentials of atoms are not well described in LSDA.\cite{Forstreuter97}
This problem is not present in bulk metals.
\item As a summary of the three points above, one can expect relativistic LSDA(+OPC) to work reasonably well for dimers, if one can assure the correct ground state configuration and ground state
spin multiplicity.
Therefore, we compare the results of scalar-relativistic calculations with available results of
other calculations and with experimental data.
\item Last but not least, we are aware of only two LSDA and/or GGA calculations\cite{Strandberg_NatureMaterials6_648,Fernandez-Seivane_prl99_183401} to study orbital
magnetic properties, in particular MAE, of dimers.
No LSDA+OPC or more involved quantum chemical calculations on this topic have been reported so far. Thus, the presented results are the first that will not only allow to estimate a lower bound (LSDA and GGA) but also an upper bound to the magnitude of MAE and trends among the considered series of transition metal dimers.
\end{itemize}

Before we proceed to more computational details, a final remark on the angle dependence of MAE is in place.
In dimers, MAE, Eq.~(\ref{EQ:MAE}), due to axial symmetry only depends on the angle $\Theta$ between the dimer axis and the magnetization. Further, extrema of MAE$(\Theta)$ are expected at $\Theta=0$ and $\Theta=\pi/2$, for symmetry reasons. Other extrema cannot be excluded but are unlikely. They could occur, e.g., due to multiple level crossing upon variation of $\Theta$. The present investigation is constraint to the evaluation of MAE$(\pi/2)$, MAE$(0)$ being zero by definition. A more detailed investigation should scan MAE$(\Theta)$ at some finer mesh.

\section{Computational details}
\label{ComputationalDetails}

We are aiming at the MAE which is due to spin-orbit coupling and hence of the order of the fine structure splitting of the molecular terms. Apart from exceptionally possible cases,
the distance in energy between molecular terms at equilibrium bond lengths (typically several 100 meV) is larger than the fine structure splitting, so one may concentrate on the MAE related to the molecular ground state term, where for the latter spin-orbit splitting has been neglected. After having identified this term and its structural properties, spin-orbit coupling will be switched on to study MAE.

Without spin-orbit coupling, a molecular term $^{2S+1} \Lambda_p$ is characterized by the spin multiplicity $2S+1$ of the total molecular spin $S$, by the absolute value of the eigenvalue $M$ of the projection of the total orbital momentum on the molecular axis ($\Lambda = \Sigma^\pm, \Pi, \Delta,\ldots$ for $|M| = 0, 1, 2, \ldots$), and by the parity $p = g,u$ with respect to inversion in electron configuration space. ($\Sigma^\pm$ are $M = 0$ states which are even/odd with respect to reflection on a plane containing the molecular axis; an $M \neq 0$ state is transformed into a $-M$ state of the same term by such a reflection.) Besides the spin multiplicity,
there is an orbital multiplicity. Within the scalar-relativistic approximation (spin-orbit coupling neglected),
spin and orbital parts of the state are only related via fermionic antisymmetry with respect to particle exchange, and the spin density matrix with respect to a chosen $z$-axis of spin may be well defined in the states of each molecular term since the Hamiltonian is spin independent. It may be chosen either even or odd with respect to an inversion through the center of the molecular geometry
(sometimes also called symmetric and symmetry-broken ground state).
In order to obtain single-determinant Kohn-Sham states representing a given term, collinear fixed spin moment (FSM) calculations\cite{Schwarz84}
are performed with initially even spin density and predefined $S_z = S$. For $S = 0$, alternatively odd spin density is enforced by initial conditions for the
self-consistency iterations.
The latter case corresponds in particular to the ground state for Cr$_2$, Mn$_2$, and Mo$_2$.
Calculations for all $S$-values are performed, and the ground state term is determined as that of lowest total energy.

These calculations are performed with the scalar-relativistic version of the full-potential local-orbital code FPLO.\cite{Koepernik_prb59_1743,FPLO}
Its newly implemented feature for calculating the
electronic structure of clusters or molecules (up to several 100 atoms) is used.
The valence basis comprised $3s3p3d\; 4s4p4d\; 5s$ states in the case of
$3d$ transition metal dimers, while a basis with the respective next higher
principle quantum number was used for the $4d$ case.
Within the LSDA,
the exchange-correlation functional by Perdew and Wang (PW92)\cite{Perdew92a} is applied.
A further trick within the FSM to improve the convergence behavior has been the variation of the temperature broadening parameter for the Brillouin zone integration (from 500 K stepwise down to 100 K), and varying the numerical procedures for the self-consistency loop within the calculations. The obtained ground state multiplicities $2S+1$ for the 3$d$ and 4$d$ transition metal dimers are given in the second column of Tabs.~\ref{StructProp3dTMDimers} and \ref{StructProp4dTMDimers}, respectively.

\begin{figure}
\includegraphics[width=0.8\textwidth,clip]{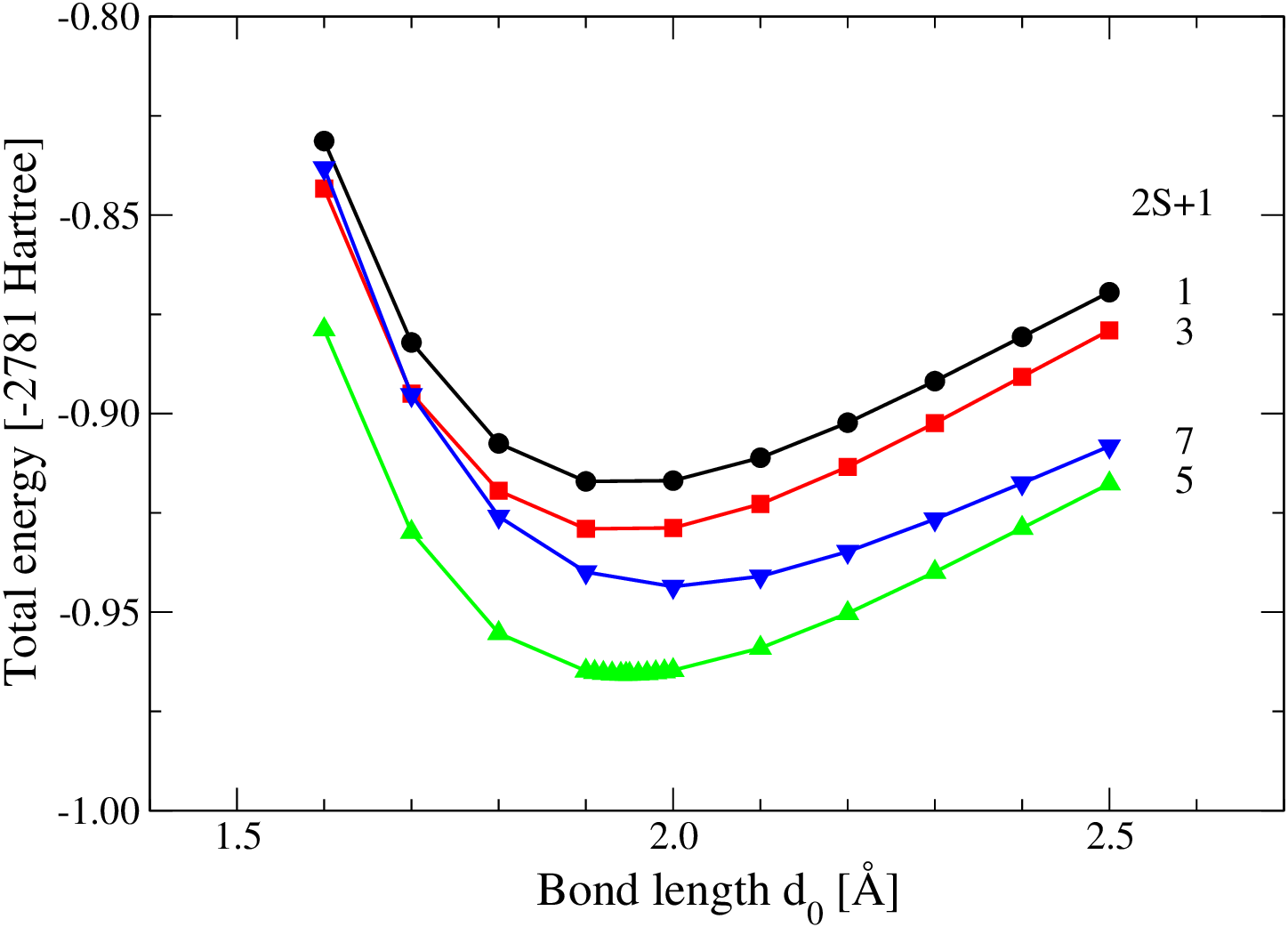} %0.45
\caption{\label{Co2_FSM}Total energy curves of the Co$_2$ dimer obtained within scalar-relativistic approximation for various spin multiplicities $2S+1$.}
\end{figure}

After having identified the ground state spin multiplicity and symmetry for each of the 3$d$ and 4$d$ transition metal dimers within our LSDA approach in scalar-relativistic approximation, we have obtained the bond lengths and harmonic vibrational frequencies by analyzing the total energy curves belonging to the dimer ground states. The necessary curve fitting has been done utilizing cubic splines. Exemplarily,
the total energy curves for various spin multiplicities $2S+1$ are given for the Co$_2$ dimer in Fig.~\ref{Co2_FSM}.

Being mainly interested in the orbital magnetic properties of the 3$d$ and 4$d$ transition metal dimers, we have then performed full-relativistic FPLO calculations\cite{FPLO,e180}
at the dimer ground state bond lengths and spin multiplicities obtained by previous calculations. It is known from earlier investigations on transition metal bulk systems, see discussion in Sec.~\ref{MagneticAnisotropy},
that treating the orbital magnetism in LSDA or GGA
leads to severe disagreement with experimental observations. Therefore, additionally orbital polarization corrections have been introduced to calculate orbital moment related properties. Within this work all the 3$d$ and 4$d$ transition metal dimers have been treated within the standard full-relativistic LSDA approach. Additionally, we performed calculations taking into account orbital polarization corrections, as described in Section \ref{OrbDepPotDFT}.

\section{Results}
\label{Results}

In this section, collecting the results of the calculations,
we first give the structural properties of the dimers 
for the ground state spin multiplicities $2S+1$
obtained within the scalar-relativistic
LSDA approach. In a second part,
all results on the magnetic properties obtained by full-relativistic
calculations are summarized.

\subsection{Spin multiplicities and structural properties}
\label{StructuralProperties}

The bond lengths and harmonic vibrational frequencies for the 3$d$ and 4$d$ transition metal dimers obtained within the scalar-relativistic LSDA approach are given for the ground state spin multiplicities in Tables \ref{StructProp3dTMDimers} and \ref{StructProp4dTMDimers}, respectively, and shall now be discussed in detail. The tables are arranged such that the first line for each dimer contains our ground state data and other data which refer to the same ground state. If other calculations or experiment yield another ground state, additional lines are added and our results for this state (being an excited state according to our calculation) are included for comparison.

\begin{figure}
\includegraphics[width=0.8\textwidth,clip]{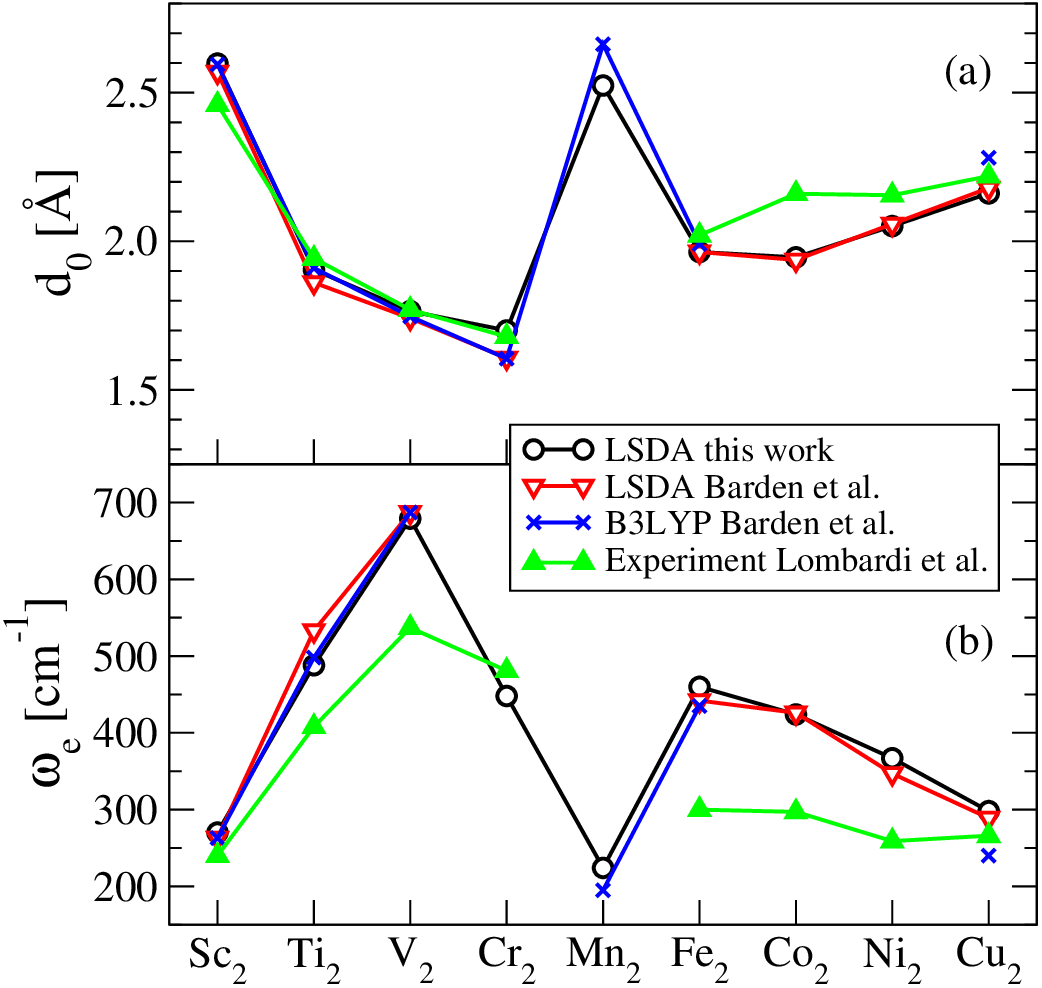} %0.475
\caption{\label{3dTMD_SP}Structural properties of the 3$d$ transition metal dimers: (a) equilibrium bond length d$_0$ [{\AA}], and (b) harmonic vibrational frequency $\omega_{e}$ [cm$^{-1}$]. Note, that some estimated values are
included for the experimental bond lengths and some data refer to ground
states of different spin multiplicities. More details are given in the related
Table \ref{StructProp3dTMDimers}.}
\end{figure}

Starting with the 3$d$ transition metal dimers, our approach predicts the ground state of the early transition metal dimers Sc$_2$, Ti$_2$, and V$_2$ to have spin multiplicity 5, 3, and 3, respectively, in agreement with other calculations and experimental observations. The bond lengths, given in Fig.~\ref{3dTMD_SP} (a), agree reasonably well (within 2\%) with other LSDA calculations performed by Barden \textit{et al.}\cite{Barden_JChemPhys113_690} An even better agreement (within 1\%) is obtained, probably by coincidence, with the B3LYP calculations by the same authors.\cite{Barden_JChemPhys113_690}
A similar good agreement is found between our calculations and experiment for Ti$_2$ and V$_2$ (note, that the value for Sc$_2$ given in column `Exp' is an estimation). The calculated vibrational frequencies agree well with each other and deviate in a systematic way from the experimental data, see Fig.~\ref{3dTMD_SP} (b).

Continuing further, the Cr$_2$ dimer shows
competition between low-lying even and odd zero-spin states.
Our calculation yields an odd ground state, in agreement with the LSDA calculation 
by Valiev \textit{et al.}\cite{Valiev_JChemPhys119_5955}
This is at variance with the results by Barden \textit{et al.}\cite{Barden_JChemPhys113_690}
who report an even ground state obtained both in LSDA and in B3LYP.
We show structural data for both states in Table~\ref{StructProp3dTMDimers}.
In our calculation,
the even state (with zero spin density) lies 127 meV above the symmetry-broken
ground state.
The respective bond lengths agree within about 2\% as for the previous
examples. Remarkably, the vibrational frequencies are considerably different
for both states, the state with a non-zero but odd spin density providing
a much softer harmonic potential than the zero spin-density state.
As the experimental frequency agrees with the calculated frequencies
belonging to the odd state, we assume that this is the true ground state
and present the experimental data in this line.
The Cr$_2$ dimer is excluded from the further investigation, since
its zero-spin ground state is in any case
hardly susceptible to moderate applied external magnetic fields.

The Mn$_2$ dimer is difficult to treat within LSDA.
This is due to the high promotion energy from the atomic
$(3d)^5(4s)^2$ configuration to the $(3d)^6(4s)^1$ configuration
preventing strong hybridization and making van der Waals interactions
important.\cite{Yamamoto_JChemPhys124_124302}
Further, several ground states with different spin multiplicities between
1 and 11 compete.
We find a ground state with the latter value
in agreement with DFT calculations performed by Gutsev 
\textit{et al.}\cite{Gutsev_JPhysChemA107_4755}
However, the experimental ground state is a singlet state,
probably with odd non-zero spin density.\cite{Lombardi_ChemRev102_2431,Yamamoto_JChemPhys124_124302}
The Mn$_2$ dimer is therefore excluded from further consideration.

For the remaining 3$d$ transition metal
dimers we obtain again good agreement with
other calculations and fair agreement with experiment.
No experimental data are known for the ground state
spin multiplicities of Fe$_2$ and Co$_2$. Here, our
results agree very well with other calculations.
For Fe$_2$, also the experimental bond length compares well
with the calculations, but for both Fe$_2$ and Co$_2$ the
vibrational frequencies are lower in the experiment than in the
DFT calculations. The same holds for Ni$_2$, where the
experimental ground state is reported to be a mixture of a
triplet and a singlet state due to spin-orbit 
coupling.\cite{Lombardi_ChemRev102_2431}
Our calculations yield a triplet ground state,
in agreement with the LSDA calculation by
Valiev \textit{et al.},\cite{Valiev_JChemPhys119_5955} but in
disagreement with the calculation by
Barden \textit{et al.}\cite{Barden_JChemPhys113_690} who find a singlet
state with even spin density. We also provide data for this state which is found at 414 meV in our calculation. The full-relativistic calculations will be carried out for the triplet state of Ni$_2$. They should be taken with caution in view of the mentioned spin-orbit mixing between singlet and triplet states.

The copper dimer has a singlet ground state with zero spin density and
is not considered further. Its structural properties obtained in the
calculations agree reasonably well with the experimental data.

\begin{figure}
\includegraphics[width=0.8\textwidth,clip]{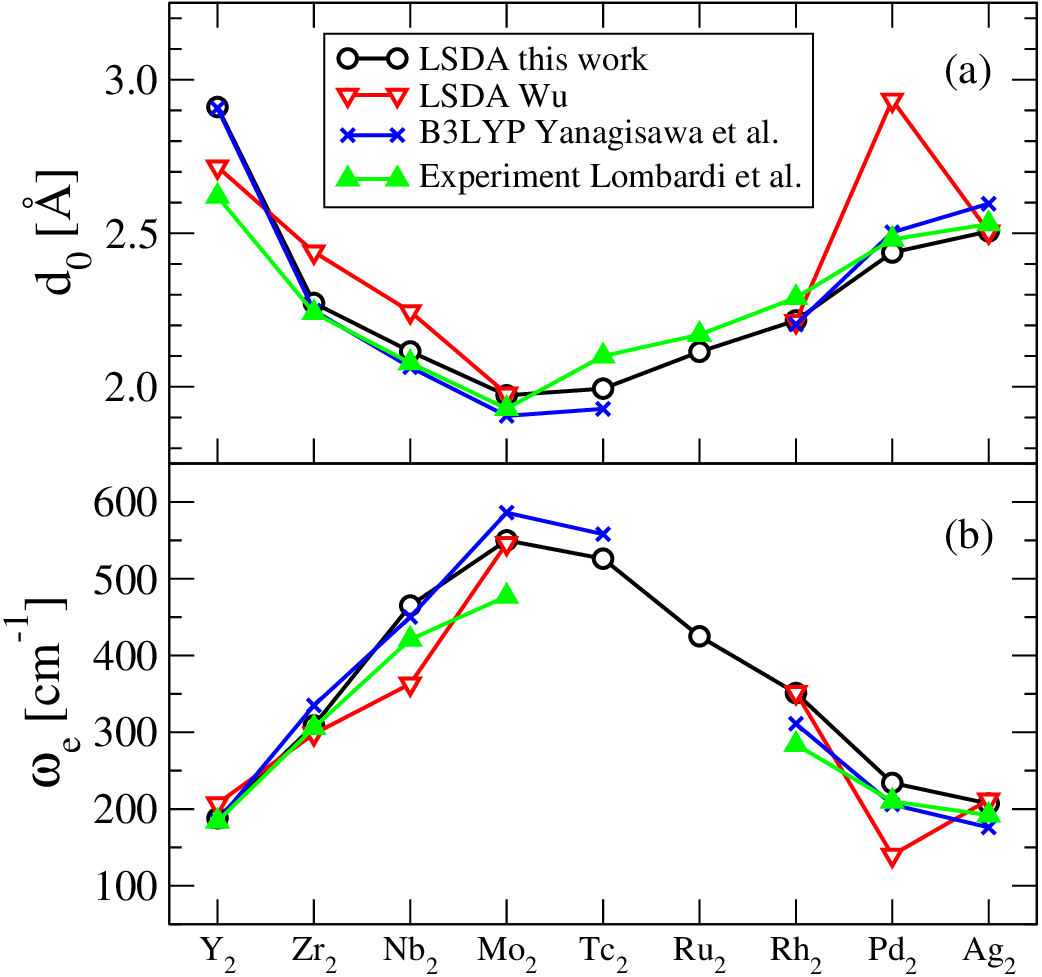} %0.475
\caption{\label{4dTMD_SP}Structural properties of the 4$d$ transition metal dimers: (a) equilibrium bond length d$_0$ [{\AA}], (b) harmonic vibrational frequency $\omega_{e}$ [cm$^{-1}$]. Note, that some estimated values are
included for the experimental bond lengths and some data refer to ground
states of different spin multiplicities. More details are given in the related
Table \ref{StructProp4dTMDimers}.}
\end{figure}

Continuing with the 4$d$ transition metal dimers, their bond lengths and harmonic vibrational frequencies are given for the respective
ground state multiplicities in Tab.~\ref{StructProp4dTMDimers}.
Quite similar to the 3$d$ transition metal dimers discussed above, the results obtained for the early 4$d$ transition metal dimers Y$_2$, Zr$_2$, and Nb$_2$ are in agreement with other theoretical calculations and experiment. The correct ground state spin multiplicities of 5, 3, and 3, respectively, are obtained for these systems in our calculations.
The bond lengths and harmonic vibrational frequencies agree with
other reported values to within a few percent, as can be seen in Fig.~\ref{4dTMD_SP} (a) and (b). The FPLO results for
the 4$d$ transition metal dimers are in a better agreement with the calculations performed by Yanagisawa \textit{et al.}\cite{Yanagisawa_JCompChem22_1995} than with the calculations performed by Wu.\cite{Wu_ChemPhysLett383_251}

Proceeding further, the Cr column transition metal dimer Mo$_2$ has been
reported to show a broken-symmetry (odd zero-spin) 
ground state in LSDA.\cite{Delley_prl50_488}
Our calculation yields a zero spin density ground state,
but the odd zero-spin ground state lies only a few meV above.
We exclude Mo$_2$ from further consideration.
Experimental and theoretical structural data show reasonable agreement,
with a somewhat smaller vibrational frequency in experiment compared
with the calculations. One is tempted to take this difference as
hint to an odd zero-spin ground state in experiment.
On the other hand, the vibrational frequency calculated for such
a state, $520 \pm 50$ cm$^{-1}$,\cite{Delley_prl50_488}
is not clearly different from 
the related frequencies calculated for even zero-spin ground state,
at variance with the situation in Cr$_2$.

For the Ru$_2$ dimer,
we obtain a ground state spin multiplicity of 5, in disagreement with other calculations which yield a ground state multiplicity of 7.\cite{Wu_ChemPhysLett383_251,Yanagisawa_JCompChem22_1995}
Our calculated state with $2S+1 = 7$ lies approximately 400 meV above the
ground state. Its structural data 
agree well with related literature values and also with the experimental
vibrational frequency. The latter supports a $2S+1 = 7$ ground state.
While we could stabilize this state in the scalar-relativistic 
calculations, we failed to achieve this in the full-relativistic
approach. This is the reason why no data for Ru$_2$ are provided in the
next section.

For the remaining 4$d$ transition metal dimers Rh$_2$, Pd$_2$, and Ag$_2$ we can adopt the discussion of the late 3$d$ transition metal dimers. The agreement with other calculations and experimental results is reasonable, and the agreement with calculations performed by Yanagisawa \textit{et al.}\cite{Yanagisawa_JCompChem22_1995} is better than with results by Wu.\cite{Wu_ChemPhysLett383_251}
The non-magnetic Ag$_2$ dimer is excluded from further consideration.

In general, we find quite good agreement for the 3$d$ and 4$d$ transition metal dimers concerning the ground state spin multiplicity, their
bond lengths and harmonic vibrational frequencies.
Ground states with wrong spin multiplicity are found for Mn$_2$ and Ru$_2$.
Comparing our results with the most recent work by 
Strandberg \textit{et al.},\cite{Strandberg_NatureMaterials6_648} 
who considered Co$_2$, Fe$_2$, Ni$_2$, Rh$_2$, Ru$_2$, and Pd$_2$,
we find the same ground state multiplicities for all systems including Ru$_2$.

\subsection{Orbital magnetic properties}
\label{MagneticProperties}

After having identified the spin multiplicities of the dimer ground states within a scalar-relativistic treatment, we now discuss the results on the orbital magnetic properties obtained by the full-relativistic approach to the dimer ground states. The spin ($\mu_s$) and orbital ($\mu_l$) moments of the 3$d$ and 4$d$ transition metal dimers are given in Tables \ref{MagMom3dTMDimers} and \ref{MagMom4dTMDimers}, respectively, for magnetizations directing perpendicular to and along the dimer axis. For each dimer the results of the full-relativistic LSDA calculations are given in the first row, whereas in the second row results are compiled obtained by full-relativistic LSDA+OPC calculations as discussed in Sec.~\ref{MagneticAnisotropy}. Generally, in small systems like dimers orbital moments are less quenched compared to systems with larger coordination numbers, and second Hund's rule couplings are more important.

The orbital moments are relatively small for the first and third transition metal dimers of the 3$d$ and 4$d$ rows. This is in agreement with the experimentally observed\cite{Lombardi_ChemRev102_2431} zero orbital angular momentum ground states. (The values of the ground state orbital angular momenta compiled by Lombardi \textit{et al.},\cite{Lombardi_ChemRev102_2431} are given in parentheses behind the easy-axis
calculated orbital moments, Tables \ref{MagMom3dTMDimers} and \ref{MagMom4dTMDimers}). Orbital polarization corrections enhance the orbital moments up to a factor of two, but the MAE is relatively small for both LSDA and LSDA+OPC approaches, as spin-orbit coupling has little influence on the $L=0$ states.

The situation is different for the Ti$_2$ and Zr$_2$ dimers with experimentally observed $L=2$ ground states. Here, we obtain significant orbital moments and anisotropy energies. As a grain of salt, the evaluated orbital moment of the Ti$_2$ dimer is smaller than expected. This might be a problem of LSDA.
It is only partly cured by the application of OPC, which changes the sign of the orbital moment along the dimer axis of Ti$_2$. This is related with a {\em reduction} of the MAE. The opposite holds for Zr$_2$, where OPC enhances the MAE to a considerable value of 0.08 eV. In this case, the calculated orbital moment is close to the experimental one. It should be noted that these experimental values exist only for the {\em easy} axis of magnetization, i.e., for the ground state orientation. The same holds for Fe$_2$ and Co$_2$, where the orbital moments along the easy axis are very close to 2 $\mu_{\rm B}$ reported in literature (calculations), but are much smaller along the hard axis. For Ni$_2$ and, partly, for Pd$_2$ the situation is less clear. This might be due to a stronger spin-orbit mixing in these systems
and possible level crossings that deserve further investigation.

Very interesting results are obtained
for the magnetic anisotropy energy of the late 
transition metal dimers beginning with Fe$_2$ and Tc$_2$ for the 3$d$
and 4$d$ row, respectively.
For the Co$_2$ and Rh$_2$ dimers we obtain an MAE of $+50$ meV (LSDA) and
$+104$ meV (LSDA), respectively, in agreement with
$+60$ meV and $+90$ meV reported by Strandberg 
\textit{et al.}\cite{Strandberg_NatureMaterials6_648} 
Due to OP corrections these values are enhanced to $+188$ meV and $+196$ meV,
respectively. 
Another large MAE value is found for Fe$_2$, almost competing in size
with the two previously mentioned dimers, Co$_2$ and Rh$_2$.
The record value of 223 meV obtained for Ni$_2$ by LSDA+OPC should be
taken with caution, since Ni$_2$ has a multiconfiguration ground state.

Intriguingly, all large values of MAE in the order of $0.1$ eV are
related with uniaxial anisotropy and anisotropy of the orbital
moment magnitude. This can be understood from the interplay
of bonding and spin-orbit splitting. If the magnetization is parallel
to the dimer axis, orbitals with zero magnetic quantum number form
$\sigma$ bonds. The states with highest magnetic quantum numbers,
$|m_l| = 2$, form $\delta$ bonds instead. Spin-orbit interaction can
gain considerable energy by splitting these states.
On the other hand, if the magnetization and, thus, the quantization
axis, is perpendicular to the dimer
axis, the $\sigma$ bonds are formed by combined states with $m_l = \pm 2$.
In this case, there is no effect of spin-orbit interaction on these
states and the energy is higher. Similar arguments have been provided by
Strandberg \textit{et al.}\cite{Strandberg_NatureMaterials6_648}

\section{Summary and Outlook}
\label{SummaryAndOutlook}

We have reported a systematic density functional study of ground state properties, in particular of the magnetic anisotropy energy, for the homonuclear 3$d$ and 4$d$ transition metal dimers. The notation of magnetic anisotropy, central to this paper but less common in chemistry, has been introduced in some detail. Before, orbital dependent potentials in density functional theory have been discussed, as a reasonable description of orbital magnetism including magnetic anisotropy requires corrections to the most common LSDA or GGA approaches within DFT. The presented results have been obtained by scalar- and full-relativistic LSDA calculations, the latter without and with orbital polarization corrections applied.
To ensure consistency of the calculations with previous work, ground state spin multiplicities, bond lengths, and vibrational frequencies have been compared with available theoretical and experimental data from literature. Good agreement between the present data and results from literature has been found for most cases. Spin and orbital moments as well as MAE have been evaluated for all dimers with $2S+1 > 1$ in the ground state, except Ru$_2$, where our approach did not find the correct ground state spin multiplicity.

The main finding of the present work is an exceptionally large magnetic anisotropy energy of about 0.2 eV/f.u. for Fe$_2$, Co$_2$, Ni$_2$, and Rh$_2$, obtained by means of LSDA+OPC. Values of 0.06 eV/f.u. and
0.09 eV/f.u. have recently been reported in literature\cite{Strandberg_NatureMaterials6_648} for Co$_2$ and Rh$_2$, respectively, obtained by means of LSDA calculations. We confirm these LSDA values but note that LSDA is known to provide, in many cases, only a lower bound on the strength of orbital magnetic properties including MAE, while LSDA+OPC frequently yields results closer to experiment.\cite{Richter98}
Another set of LSDA and GGA calculations on dimer MAE was reported very recently, where large numbers have been found for the MAE of Pt$_2$ (0.035 eV with LSDA and 0.1 eV with GGA) and of Ir$_2$ (0.06 eV with LSDA).\cite{Fernandez-Seivane_prl99_183401} The quoted paper also reports an LSDA value, -0.005 eV, for the MAE of Pd$_2$, in agreement with our related value. Finally, a small negative MAE is found for Ni$_2$ in Ref. \onlinecite{Strandberg_NatureMaterials6_648}, in disagreement with our positive value of 0.011 eV. We attribute this difference to the mentioned problem to describe the ground state of Ni$_2$ with LSDA-like approaches. A full-relativistic many-electron treatment of such systems would be desirable but seems hard to perform at present.

All these data raise expectations about the potential applicability
of magnetic dimers for information storage.
In particular, a barrier height of 0.2 eV as obtained in the present
study would translate into
10 years stability of superparamagnetic bits at a temperature of 60 K.
It is therefore tempting to search for other homo- or heteronuclear
dimers with yet higher MAE.
A further necessary and perhaps not unrealistic
step would be to find a matrix that can fix the dimers
in space without deteriorating their magnetic properties.
Finally and probably most difficult, a completely new read/write technology
would have to be invented for any practical application of such small
magnetic entities.

Besides these dreams, exceptional magnetic properties have always
been a playground
for fundamental research. Magnetic dimers are potential model systems
in this field due to their unique combination of structural simplicity and
large magnetic anisotropy.

\section{Acknowledgements}
\label{Acknowledgements}

We gratefully acknowledge illuminating discussions with Michael Kuz'min, Gotthard Seifert, and Cyrille Barreteau. Financial support was provided by Deutsche Forschungsgemeinschaft within SPP 1145 (grant ES 85/10-3).

%\bibliographystyle{apsrev}
%\bibliography{JCompChem}

\cleardoublepage %Figures

\newpage

\newpage

\newpage %Tables

\begin{table*}
\caption{\label{StructProp3dTMDimers}
Spin multiplicity $2S+1$, bond length d$_0$ [{\AA}], and harmonic vibrational frequency $\omega_e$ [cm$^{-1}$] for the homonuclear 3$d$ transition metal dimers.
Results from this work (abbreviated TW) are compared with experiment and
with other DFT calculations from literature.
Zero-spin states with even (zero) spin density are denoted by (e) and
those with odd (non-zero) spin density by (o).}
\begin{ruledtabular}
\begin{tabular}{ccccccccccc}
Dimer  & \multicolumn{2}{c}{$2S+1$} & \multicolumn{4}{c}{d$_0$ [{\AA}]} & \multicolumn{4}{c}{$\omega_{e}$ [cm$^{-1}$]} \\
\cline{2-3}
\cline{4-7}
\cline{8-11}
       & TW & Exp\footnotemark[1]   & TW & LSDA\footnotemark[2] & B3LYP\footnotemark[3] & Exp\footnotemark[1]
            & TW & LSDA\footnotemark[2] & B3LYP\footnotemark[3] & Exp\footnotemark[1] \\
\hline
Sc$_2$ & 5 & 5 & 2.598 & 2.569 & 2.595 & (2.46)\footnotemark[4]  & 270 & 263 & 263 & 240 \\
Ti$_2$ & 3 & 3 & 1.904 & 1.861 & 1.910 & 1.942 & 488 & 533 & 498 & 408 \\
V$_2$  & 3 & 3 & 1.765 & 1.741 & 1.747 & 1.77  & 679 & 687 & 687 & 537 \\
Cr$_2$ &  1(o) & 1 & 1.700 & 1.66\footnotemark[5]  &       & 1.679 & 448 & 557\footnotemark[5] &     & 481 \\
       &  1(e) & & 1.626 & 1.607 & 1.605 &       & 799 & 821 & 837 & \\
Mn$_2$ & 11 &   & 2.524 &       & 2.663 &       & 224 &     & 195 & \\
       & 1(o) & 1 & 2.429 &       &       & 3.4   & 144 &     &     & 76.4 \\
Fe$_2$ &  7 & \{7\}\footnotemark[6] & 1.964 & 1.964 & 1.991 & 2.02  & 460 & 442 & 435 & 300 \\
Co$_2$ &  5 & \{5\}\footnotemark[6] & 1.946 & 1.937 &       & (2.16)\footnotemark[4]  & 424 & 426 &     & 297 \\
Ni$_2$ &  3 & 3\footnotemark[7] & 2.051 & 2.054\footnotemark[5] &       & 2.154 & 367 & 328\footnotemark[5] &     &  259    \\
       &  1(e) & & 2.054 & 2.058 & 2.296 &       & 360 & 347 & 227 & \\
Cu$_2$ &  1(e) & 1 & 2.161 & 2.179 & 2.281 & 2.219 & 298 & 289 & 240 & 266 \\
\end{tabular}
\footnotetext[1]{Experimental results taken from Lombardi and Davis.\cite{Lombardi_ChemRev102_2431}}

\footnotetext[2]{LSDA calculations by Barden \textit{et al.}\cite{Barden_JChemPhys113_690}}

\footnotetext[3]{B3LYP calculations by Barden \textit{et al.}\cite{Barden_JChemPhys113_690}}

\footnotetext[4]{Estimated in Ref.~\onlinecite{Lombardi_ChemRev102_2431} using Pauling's relation.}

\footnotetext[5]{Taken from calculations by Valiev \textit{et al.}\cite{Valiev_JChemPhys119_5955}}

\footnotetext[6]{Suggested from density functional calculations quoted in
Ref.~\onlinecite{Lombardi_ChemRev102_2431}.}

\footnotetext[7]{Due to strong spin-orbit coupling, the ground state is
a mixture of states with spin multiplicities 3 and 1.\cite{Lombardi_ChemRev102_2431}}
\end{ruledtabular}
\end{table*}

\newpage

\begin{table*}
\caption{\label{StructProp4dTMDimers}
Spin multiplicity $2S+1$, bond length d$_0$ [{\AA}], and harmonic vibrational frequency $\omega_e$ [cm$^{-1}$] for the homonuclear 4$d$ transition metal dimers.
Results from this work (abbreviated TW) are compared with experiment and
with other DFT calculations from literature.
Zero-spin states with even (zero) spin density are denoted by (e) and
those with odd (non-zero) spin density by (o).}
\begin{ruledtabular}
\begin{tabular}{ccccccccccc}
Dimer  & \multicolumn{2}{c}{$2S+1$} & \multicolumn{4}{c}{d$_0$ [{\AA}]} & \multicolumn{4}{c}{$\omega_{e}$ [cm$^{-1}$]} \\
\cline{2-3}
\cline{4-7}
\cline{8-11}
       & TW & Exp\footnotemark[1] & TW & LSDA\footnotemark[2] & B3LYP\footnotemark[3] & Exp\footnotemark[1]
            & TW & LSDA\footnotemark[2] & B3LYP\footnotemark[3] & Exp\footnotemark[1] \\
\hline
Y$_2$  &  5 & 5 & 2.911 & 2.716 & 2.907 & (2.62)\footnotemark[4]  & 188 & 207 & 184 & 184 \\
Zr$_2$ &  3 & 3 & 2.273 & 2.439 & 2.248 & 2.241 & 309 & 298 & 335 & 306 \\
Nb$_2$ &  3 & 3 & 2.115 & 2.245 & 2.064 & 2.078 & 465 & 363 & 450 & 425 \\
Mo$_2$ &  1(e) & 1 & 1.972 & 1.977 & 1.976\footnotemark[5] & 1.929 & 550 & 546 & 548\footnotemark[5] & 477 \\
Tc$_2$ &  3 & \{3\}\footnotemark[3] & 1.994 &       & 1.928 & (2.10)\footnotemark[4]  & 526 &     & 558 &   \\
Ru$_2$ &  5 &  & 2.114 &       &       &       & 425 &     &     &     \\
       & & \{7\}\footnotemark[6] & 2.238 & 2.236 & 2.274 & (2.17)\footnotemark[4]  & 348 & 327 & 306 & 347 \\
Rh$_2$ &  5 & \{5\}\footnotemark[6] & 2.216 & 2.213 & 2.203 & (2.29)\footnotemark[4]  & 351 & 352 & 311 & 284 \\
Pd$_2$ &  3 & 3 & 2.437 & 2.934 & 2.503 & (2.48)\footnotemark[4]  & 234 & 140 & 206 & 210 \\
Ag$_2$ &  1(e) & 1 & 2.506 & 2.505 & 2.596 & 2.530 & 207 & 212 & 176 & 192 \\
\end{tabular}
\footnotetext[1]{Experimental results taken from Lombardi and Davis.\cite{Lombardi_ChemRev102_2431}}

\footnotetext[2]{LSDA calculations by Wu.\cite{Wu_ChemPhysLett383_251}}

\footnotetext[3]{B3LYP calculations by Yanagisawa \textit{et al.}\cite{Yanagisawa_JCompChem22_1995}}

\footnotetext[4]{Estimated in Ref.~\onlinecite{Lombardi_ChemRev102_2431} using Pauling's relation.}

\footnotetext[5]{Taken from calculations by Wu.\cite{Wu_ChemPhysLett383_251}}

\footnotetext[6]{Suggested from quantum chemical calculations quoted in
Ref.~\onlinecite{Lombardi_ChemRev102_2431}.}
\end{ruledtabular}
\end{table*}

\newpage

\begin{table*}
\caption{\label{MagMom3dTMDimers}
Spin $\mu_{s}$ and orbital $\mu_{l}$ moments
of the 3$d$ transition metal dimers obtained by full-relativistic calculations
with magnetic moment orientation perpendicular and parallel to the
dimer axis, respectively.
The different values obtained by LSDA calculations and by LSDA+OPC
calculations as discussed in Sec.~\ref{MagneticAnisotropy} are given
in the first and second row for each of the dimers, respectively.
MAE$(\pi/2)$ is the magnetic anisotropy energy whose sign is defined
in Eq.~\eqref{EQ:MAE}, MAE$(\pi/2) = E_0(\perp) - E_0(\parallel)$, i.e.,
positive values correspond to an easy axis parallel to the dimer axis.
All values are given per dimer. Moments deviating less than 0.01 $\mu_{\rm B}$
from an integer value are rounded to the next integer.
The values in parentheses given behind the easy-axis orbital moments
are experimental \{theoretical\} values of the corresponding ground
state orbital angular momentum given in Table 2 of 
Ref.~\onlinecite{Lombardi_ChemRev102_2431}.}
\begin{ruledtabular}
\begin{tabular}{cccccc}
& \multicolumn{2}{c}{$\perp$} & \multicolumn{2}{c}{$\parallel$} \\
\cline{2-3}
\cline{4-5}
Dimer & $\mu_{s}$   & $\mu_{l}$   & $\mu_{s}$   & $\mu_{l}$   & MAE$(\pi/2)$ \\
      & [$\mu_{B}$] & [$\mu_{B}$] & [$\mu_{B}$] & [$\mu_{B}$] & [meV]     \\
\hline
Sc$_{2}$ & 4. & -0.040 (0)  & 4.&  0.&    - 0 \\
         & 4. & -0.062 (0)  & 4.&  0.&    - 0 \\
Ti$_{2}$ & 2. & -0.016  & 2.& -0.264 (2) &   + 36 \\
         & 2. & -0.026  & 2.&  0.434 (2) &   + 33 \\
V$_{2}$  & 2. & -0.016 (0) & 2.& 0.&    - 1 \\
         & 2. & -0.030 (0) & 2.& 0.&    - 1 \\
Fe$_{2}$ & 6. &  0.186 &  6. &  1.890 (\{2\}) &   + 32 \\
         & 6. &  0.402 &  6. &  1.908 (\{2\}) &  + 150 \\
Co$_{2}$ & 4.144 &  0.334 &  4.088 &  2. (\{2\}) &   + 50 \\
         & 4.142 &  1.196 &  4. &  2.096 (\{2\}) &  + 188 \\
Ni$_2$   & 1.986 &  0.454 &  1.988 &  0.878 &   + 11 \\%& \\
         & 1.990 &  1.468 &  2. &  3.120 &  + 223 \\%& \\
\end{tabular}
\end{ruledtabular}
\end{table*}

\newpage

\begin{table*}
\caption{\label{MagMom4dTMDimers}
Spin $\mu_{s}$ and orbital $\mu_{l}$ moments
of the 4$d$ transition metal dimers obtained by full-relativistic calculations
with magnetic moment orientation perpendicular and parallel to the
dimer axis, respectively.
The different values obtained by LSDA calculations and by LSDA+OPC
calculations as discussed in Sec.~\ref{MagneticAnisotropy} are given
in the first and second row for each of the dimers, respectively.
MAE$(\pi/2)$ is the magnetic anisotropy energy whose sign is defined
in Eq.~\eqref{EQ:MAE}, MAE$(\pi/2) = E_0(\perp) - E_0(\parallel)$, i.e.,
positive values correspond to an easy axis parallel to the dimer axis.
All values are given per dimer. Moments deviating less than 0.01 $\mu_{\rm B}$
from an integer value are rounded to the next integer.
The values in parentheses given behind the easy-axis orbital moments
are experimental \{theoretical\} values of the corresponding ground
state orbital angular momentum given in Table 2 of
Ref.~\onlinecite{Lombardi_ChemRev102_2431}.}
\begin{ruledtabular}
\begin{tabular}{cccccc}
& \multicolumn{2}{c}{$\perp$} & \multicolumn{2}{c}{$\parallel$} \\%& \\
\cline{2-3}
\cline{4-5}
Dimer & $\mu_{s}$   & $\mu_{l}$   & $\mu_{s}$   & $\mu_{l}$   & MAE$(\pi/2)$ \\
      & [$\mu_{B}$] & [$\mu_{B}$] & [$\mu_{B}$] & [$\mu_{B}$] & [meV]     \\
\hline
Y$_{2}$  & 4. & -0.104 (0) & 4. &  0.&    - 1 \\
         & 4. & -0.146 (0) & 4. &  0.&    - 1 \\
Zr$_{2}$ & 1.984 & -0.062 & 2.& -1.710 (2) &   + 25 \\
         & 1.984 & -0.116 & 2.& -1.744 (2) &   + 76 \\
Nb$_{2}$ & 1.972 & -0.044 (0) & 2.& 0.&    - 9 \\
         & 1.972 & -0.080 (0) & 2.& 0.&   - 10 \\
Tc$_{2}$ & 1.924 &  0.052 (?) & 2.&  0.012 &   - 31 \\
         & 1.924 &  0.084 (?) & 2.&  0.012 &   - 34 \\
Rh$_{2}$ & 3.926 &  0.630 & 3.978 &  2.116 (\{2\}) &  + 104 \\
         & 3.928 &  1.088 & 3.976 &  2.122 (\{2\}) &  + 196 \\
Pd$_{2}$ & 1.980 &  0.532 (0) & 1.938 &  0.862 &    - 5 \\
         & 1.984 &  0.762 & 1.940 &  1.106 (0) &   + 10 \\
\end{tabular}
\end{ruledtabular}
\end{table*}

\end{document}